\documentstyle[12pt,psfig]{article}
\begin{document}
\centerline{\Large \bf Surface Instability in Windblown Sand}
\vskip 1.0 true cm
\centerline{\bf Douglas A. Kurtze}
\vskip 0.2 true cm
\centerline{Department of Physics, North Dakota State University,
Fargo, ND  58105-5566, USA}
\vskip 0.2 true cm
\centerline {{\bf Joseph A. Both} and {\bf Daniel C. Hong}}
\vskip 0.2 true cm
\centerline{Department of Physics, Lewis Laboratory, Lehigh
University, Bethlehem, PA  18015, USA}



\date{\today}


\begin{abstract}
We investigate the formation of ripples on the surface of windblown
sand based on the one-dimensional model of Nishimori and Ouchi
[Phys. Rev. Lett. {\bf 71}, 197 (1993)], which contains the processes of
saltation and grain relaxation.  We carry out a nonlinear analysis
to determine the propagation speed of the restabilized ripple patterns,
and the amplitudes and phases of their first, second, and third harmonics.
The agreement between the theory and our numerical simulations is excellent
near the onset of the instability.  We also determine the Eckhaus boundary,
outside which the steady ripple patterns are unstable.
\end{abstract}


\section{Introduction}

Since the pioneering work of Bagnold [1], many researchers have
investigated the complex dynamics of dry granular materials at a surface
[2-6].  Dry granular materials are assemblies of macroscopic objects that
interact with each other essentially via a hard core repulsive potential.
Hence they are loosely connected, particularly at the surface.  When
those grains at the surface are exposed to a wind, they can readily be
ejected and carried by the wind until gravity eventually pulls them back
to the surface.  The dynamics of a single grain is rather simple, given
by the Newtonian trajectory of a point particle.  Even so, experiments
have shown that the collective response of the grains can become
exceedingly complex, ranging from formation of simple ripple patterns to
ridges and dunes to violent tornadoes [2].   Our current understanding of
such complex phenomena remains mostly confined to compiling data on
experimental observations.  With regard to the formation of ripple
patterns, however, there have been some attempts to construct a simple yet
physical continuum model.

We will investigate the continuum model due to Nishimori and Ouchi [3].
The Nishimori-Ouchi (NO) model of ripple patterns accounts for two
elementary processes of sand transportation by the wind which have been
identified by investigators in aeolian sand dynamics, namely saltation and
creep [1,2].  {\it Saltation} refers to the process by which surface grains
are ejected into the air under the influence of a strong wind, and are blown
downwind where they collide with other surface grains.  There they transfer
momentum to these downwind grains, which may themselves be ejected in turn,
thereby continuing the process (see Fig. 1).  {\it Creep} is the surface
movement of grains too heavy to be ejected into the air but light enough to
be pushed along the surface.  Creep also describes the surface movement of
grains on hills under the influence of gravity.  Previous studies based on
the NO model have been confined largely to linear stability analysis and
Monte Carlo simulations of a lattice version with simple rules for the grain
dynamics [3,11].  The purpose of this paper is to go one step farther by
carrying out a nonlinear analysis of the continuum model and uncovering some
of the features of ripple formation that are inaccessible to linear analysis.  

In particular, we carry out a weakly nonlinear analysis valid near the
onset of instability of a flat sandbed to determine the amplitude, shape,
and propagation speed of the ripple pattern that forms in this regime.
We also compare these results with our numerical integrations of the model
equations.  These computations are rather unusual because the model
lacks an up-down symmetry, and especially because accounting for saltation
makes the model nonlocal in space.  We find, however, that the process of
pattern selection in this simple one-dimensional system, in particular the
selection of the wavelength and speed of the patterns [7], is similar to
what is seen in more complicated multidimensional systems such as
directional solidification [8] or directional viscous fingering [9].  That
is, the wavelength of the final pattern depends on the initial conditions,
and may lie anywhere within a band of linearly stable final states.  The
stable band turns out to be somewhat wider than in most other models.

In the next section we review the Nishimori-Ouchi model equations [3],
point out a physical symmetry which they violate, and propose a simple
modification of the model which respects that symmetry.  In Section III
we carry out a linear stability analysis of the flat-sandbed solution
of the model, both for the original Nishimori-Ouchi equations and for
our modification.  We extend this in Section IV to give a weakly nonlinear
analysis for both forms of the model.  Section V presents our numerical
calculations and compares them with the results of the weakly nonlinear
analysis.  The results are discussed in the final section.

\section{One-Dimensional Model for Windblown Sand}

The starting point of the Nishimori-Ouchi (NO) model [3] is a local 
conservation law for sand grains.  Let $h(x,t)$ be the local height of the 
sand bed at position $x$ and time $t$, measured from some reference level.
The height increases when grains are added at position $x$.  We write
\begin{equation}
  {\partial h \over \partial t} + {\partial J_l \over \partial x} = Q_{nl},
\end{equation}
where $J_l(x,t)$ is a local flux of grains in the positive direction at $x$,
and $Q_{nl}(x,t)$ is the net input of grains at $x$ due to nonlocal
processes.

The expression for $J_l$ embodies a model of creep.  Nishimori and Ouchi
choose $J_l = -D(\partial h/\partial x)$.  Note that this merely expresses
the tendency of grains to roll downhill; it does not include any bias
favoring motion in the direction of the wind.  

Saltation is modeled by gain and loss terms in the nonlocal transfer rate
$Q_{nl}$.  Let $N(x,t)$ denote the outward saltation flux of particles
from $x$ at time $t$.  That is, let $N(x,t)\,dx$ be the number of
particles per unit time taking off from positions between $x$ and $x+dx$.
The loss term in $Q_{nl}$ is then $-A\,N(x,t)$, where $A$ is a scale
parameter.  The gain term is proportional to the rate at which particles
arrive at $x$ from other locations $\xi$ upwind of $x$.  Suppose all
particles which take off from the interval $(\xi,\xi+d\xi)$ subsequently
land in the interval $(x,x+dx)$.  Then the number of particles per unit
time landing in this latter interval of length $dx$ is $N(\xi,t)\,d\xi$,
so the gain term in the saltation flux is then $A\,N(\xi,t)\,(d\xi/dx)$.
It is possible to have more than one $\xi$ which satisfies this equation
for a given $x$.  That is, grains landing at $x$ may have come from more
than one takeoff point $\xi$.  If this is the case, then the input term
in the evolution equation should be summed over the different values of
$\xi$.

Note that evaluating $N(\xi)$ at time $t$ neglects the flight time of
the incoming grains; we expect the evolution of the sandbed profile to
take place on a much longer time scale than this, so that the time delay
between takeoff and landing should be unimportant. Indeed, experiments
on ripple formation by sand transported by water [10] show the evolution
of the ripple pattern occurring on time scales of several hours.

Combining the various contributions to the flux and substituting into the
general conservation law for $h$ gives the model evolution equation for
the sandbed profile,
\begin{equation}
  \frac{\partial h}{\partial t} =
  \frac{\partial}{\partial x} \, D \, \frac{\partial h}{\partial x} +
  A\,\left[ N(\xi,t) \frac{d\xi}{dx} - N(x,t) \right].
\label{conslaw1}
\end{equation}
Note that this equation is nonlocal in $x$, as a result of the saltation
gain term, which depends on conditions at a position $\xi$ which is a
finite distance upwind of $x$.

To complete the model, we must now specify the {\it saltation function},
an equation for the flight length of a single grain.  In general, the
locations $x$ and $\xi$ in the evolution equation will be related by
$x = \xi + L$, where $L$ is the horizontal length an ejected grain travels
from takeoff to landing.  This will depend on the size of the grain, its
speed when it takes off, the wind velocity profile, and the topography of
the sandbed itself.  Nishimori and Ouchi proposed the simple ansatz
\begin{equation}
  L = L_0 + b \, h(\xi,t).
\label{NO}
\end{equation}
Here, $L_0$ is a parameter proportional to the shear stress of the wind at
the surface, or more precisely to the friction velocity of the wind on the
sand surface [11], and $b$ in general depends on the average drag force on
the grain.  Nishimori and Ouchi took both $L_0$ and $b$ to be constant,
essentially assuming the wind velocity to be a constant, independent of $x$ 
and $t$ and unaffected by changes in the sandbed profile.

Equation (\ref{NO}) merely indicates that the higher the takeoff point 
of a grain in saltation, the longer its trajectory.  As Nishimori and Ouchi
point out [3], this amounts to assuming that the height and topography at
the point of landing may be neglected, and that only the surface height
(as opposed to local topography) is important at the takeoff point.  While
this may be reasonable if $h(x)$ is everywhere close to zero, it does
violate a symmetry of the physical problem, namely that the dynamics should
be unaffected if we add any constant to $h$, thus changing our reference
level.  To restore this symmetry, it may be more appropriate to take the
saltation function to be
\begin{equation}
  L = L_0 + b\,[h(\xi) - h(x)],
\label{KBH}
\end{equation}
where $\xi$ is the takeoff point and $x$ is the landing point.  We will
discuss the effects of this modification below.

For convenience, we now put the model into dimensionless form.  Taking
$L_0$, $b$ and $D$ to be constants, we choose $L_0$ to be the unit of
horizontal length, $L_0/b$ to be the unit of vertical length (i.e., of $h$),
and $L_0^2/D$ to be the time unit.  Further, we define
$J(x,t) = (AbL_0/D)\,N(x,t)$, a dimensionless measure of the outward grain
flux due to saltation.  With these definitions, the evolution equation
(\ref{conslaw1}) becomes
\begin{equation}
  \frac{\partial h}{\partial t} = \frac{\partial^2 h}{\partial x^2} + 
     J(\xi,t)\frac{d\xi}{dx} - J(x,t),
\end{equation}
with the original NO saltation relation becoming the condition
\begin{equation}
  x = \xi + 1 + h(\xi,t).
\label{NOxi}
\end{equation}

The model simplifies further if we choose $J(x,t)$ to be a constant $J$ 
independent of $x$ and $t$, an assumption whose physical content is that
the wind is uniform and there is no flux dependence on surface height.
The evolution equation is then
\begin{equation}
  \frac{\partial h}{\partial t} = \frac{\partial^2 h}{\partial x^2} + 
     J\,\left(\frac{d\xi}{dx} - 1\right).
\label{heqn}
\end{equation}
This is the form of the problem which we will analyze below, using both
the NO saltation relation (\ref{NOxi}) and our symmetric modification of it,
\begin{equation}
  x = \xi + 1 + h(\xi,t) - h(x,t).
\label{KBHxi}
\end{equation}

\section{Linear Stability Analysis}

We first note that a flat sandbed, $h = h_0 = constant$, is always a 
steady-state solution of the model, for either choice of saltation
relation.  For the symmetric saltation relation (\ref{KBHxi}) this always
gives $\xi = x - 1$, while for the NO relation (\ref{NOxi}) we have
$\xi = x - 1 - h_0$.  In the latter case, however, we may then redefine
the length and time units -- and the value of $J$ -- to map the solution
with any finite $h_0$ (provided $h_0 > -1$) onto the solution with
$h_0 = 0$.  Specifically, we would take the horizontal length unit to be
$L_0 (1+h_0)$ instead of $L_0$, and $J$ would then be $A b L_0 (1+h_0) N / D$
rather than $A b L_0 N / D$.  Thus we will take $h = 0$ to be the steady
state whose stability we will investigate.

When $h$ is small, we may linearize the NO saltation relation to get
\begin{equation}
  \xi \approx x - 1 - h(x-1),
\end{equation}
so that there is a single, unique $\xi$ for each $x$.  From this we obtain
$d\xi/dx = 1 - h'(x-1)$, where the prime indicates partial differentiation
with respect to $x$.  The linearized evolution equation is then
\begin{equation}
  \frac{\partial h}{\partial t} = h''(x,t) - J\,h'(x-1,t).
\end{equation}
Linear stability analysis proceeds in the usual way:  we write $h(x,t)$ as
a linear combination of Fourier modes,
\begin{equation}
  h(x,t) = \int h_k(t) \, \exp(ikx) \, dk,
\end{equation}
substitute this into the linearized evolution equation, and note that --
even with the nonlocal term present -- the modes do not couple.  Thus we
find that each mode grows or decays exponentially with time,
\begin{equation}
  h_k(t) \propto \exp[(\sigma_k - i\omega_k)t],
\end{equation}
with [3,11]
\begin{eqnarray}
  \sigma_k &=& - k^2 - J \, k \sin k
\label{sigmaeqn} \\
  \omega_k &=& J \, k \cos k.
\label{omegaeqn}
\end{eqnarray}
If we use the symmetric saltation relation instead of the NO relation,
then the linearized evolution equation has an extra term $+J \, h'(x,t)$
on the right side.  This leaves $\sigma_k$ unchanged, but replaces
(\ref{omegaeqn}) by $\omega_k = - J \, k \, (1 - \cos k)$.

The locus $\sigma_k = 0$ in the $J$--$k$ plane defines a stability boundary.
Points on one side of the boundary represent perturbations which have a
positive growth rate $\sigma_k$, while points on the other side represent
perturbations with negative growth rates which are therefore suppressed in
the solution.  Thus we expect that solutions of the full differential
equation will consist only of modes whose wave numbers are on the unstable
side of the boundary.  The onset of instability of the flat sandbed
occurs at the value $J_c$ of $J$ for which only a single mode, with
wave number $k_c$, is marginally stable and no other modes are unstable.
These critical values may be determined by solving $\sigma = 0$ and
$d\sigma/dk = 0$ simultaneously, which yields
\begin{eqnarray}
  J_c \, \sin k_c &=& -k_c, \nonumber \\
  J_c \, \cos k_c &=& -1.
\end{eqnarray}
Eliminating $J_c$ gives
\begin{equation}
  \tan k_c = k_c.
\end{equation}
Thus the critical values are computed as $k_c = 4.493$ and $J_c = 4.603$.
The wavelength of the marginal mode (in units of $L_0$) is
$\lambda_c = 2\pi/k_c = 1.398$, somewhat longer than the flight distance
of a grain in saltation.
For $k = k_c$, the NO saltation relation leads to $\omega_c = -k_c$, so
the phase velocity of the marginal mode is $v = \omega_c/k_c = -1$.  With
the symmetric saltation relation we get $v = -(1+J_c) = -5.60$.  This is
a surprising result of the model, that while the sand grains that form the
ripples are blown downwind, the ripple pattern itself drifts {\it upwind}.
The group velocity, however, is large and positive:  From (\ref{omegaeqn})
we get $d\omega_k/dk = J(-k\sin k + \cos k)$, which goes to $k_c^2 - 1 =
19.19$ at the critical point.  For the symmetric saltation relation, the
group velocity is lower by $J$, so at critical it is 14.58.  Note that all
velocities are in units of $D/L_0$.

If we make the problem two-dimensional, allowing the sandbed to extend in
both $x$ and $y$ directions, very little changes.  The creep term in the
evolution equation becomes $D\,\nabla^2 h$, and as a result the expression
for $\sigma_k$ changes to
\begin{equation}
  \sigma(k,k_y) = -k^2 - k_y^2 - J \, k \sin k,
\end{equation}
where $k$ is now the $x$ component of the wave vector of the Fourier mode
and $k_y$ is its $y$ component.  Clearly, the linear growth rate for a mode
with nonzero $k_y$ is always less than the rate for the corresponding mode
with $k_y = 0$.  Thus we do not expect to see instabilities in which the
transverse shape of the ripples becomes wavy, since the first instability
to occur is against a mode in which the ripples are parallel to the $y$
axis.

\section{Nonlinear Analysis}

We now carry out a weakly nonlinear analysis to determine the amplitude,
shape, and propagation velocity of the restabilized ripple patterns which
form when $J$ is slightly above its critical value $J_c$.  The nonlocality
of the model, the dispersion in the imaginary part of the linear growth
rate, and the lack of an up-down symmetry lead to some unusual features
in the analysis.

We begin with the assumption that the fundamental wave number $k$ of the
pattern which develops does not deviate much from the critical value $k_c$
when $J$ is near $J_c$.  Hence we define a small parameter $\epsilon$ by
setting
\begin{equation}
  J = J_c + \epsilon^2,
\label{Jexpansion}
\end{equation}
and then define a scaled wave number deviation $q$ by writing
\begin{equation}
  k = k_c + {\epsilon}q.
\label{kexpansion}
\end{equation}
This is the appropriate scaling for the wave number because the stability
boundary is approximately quadratic in $k-k_c$ and linear in $J$ near its
maximum.  Substituting these expressions into the linear growth rate
(\ref{sigmaeqn}) and expanding to second order in $\epsilon$ gives
\begin{equation}
  \sigma_k = {1 \over 2} \epsilon^2 k_c^2
             \left( {2 \over J_c} - q^2 \right) + O(\epsilon^3).
\label{wnlsigma}
\end{equation}
Furthermore, from the expression (\ref{omegaeqn}) for $\omega$ we find
that the phase velocity of the ripples is given by
\begin{equation}
  v = \omega/k = -1 + \epsilon k_c q -
       {1 \over 2} \epsilon^2 \left( {2 \over J_c} - q^2 \right)
       + O(\epsilon^3).
\label{wnlv}
\end{equation}

The first stage of the nonlinear analysis consists of expanding the
evolution equation (\ref{heqn}) in powers of $h$, assuming the overall
amplitude of $h$ is small.  To do this, we may rewrite the NO saltation
relation (\ref{NOxi}) in the form
\begin{equation}
  \xi = x - 1 - h(\xi,t),
\end{equation}
repeatedly substitute this expression for $\xi$ back into the $h(\xi,t)$
on the right side, and finally expand in powers of $h$.  Differentiating
the result with respect to $x$ then gives
\begin{equation}
  {d\xi \over dx} = 1 - h'(x-1,t) + {1 \over 2} [h^2(x-1,t)]''
                    - {1 \over 6} [h^3(x-1,t)]''' + O(h^4).
\end{equation}
To third order in $h$, then, the evolution equation becomes
\begin{equation}
  {\partial h(x,t) \over \partial t} = h''(x,t) - J h'(x-1,t) + 
  {J \over 2} [h^2(x-1,t)]'' - {J \over 6} [h^3(x-1,t)]''' + \cdots
\label{wnleqn}
\end{equation}
This is the equation whose ripple solutions we will presently compute.

It is remarkable that the expansion of $d\xi/dx$ has such an economical
form.  In fact it is not difficult to show that the pattern continues to
all orders in $h$.  To see this, consider the integral
\begin{equation}
  I_f \equiv \int_{-\infty}^\infty f(x-1) \, {d\xi \over dx} \, dx,
\end{equation}
where $\xi(x)$ is given by the saltation relation (\ref{NOxi}) and $f$ is
a test function which is integrable and infinitely differentiable, but
otherwise arbitrary.  We now change variables in this integral from $x$
to $\xi$,
\begin{equation}
  I_f = \int_{-\infty}^\infty f(\xi+h(\xi)) \, d\xi,
\end{equation}
and expand the integrand in powers of $h$ to get
\begin{equation}
  I_f = \int \sum_{k=0}^\infty {1 \over k!} \, {d^k f(\xi) \over d\xi^k}
        \, h^k(\xi) \, d\xi.
\end{equation}
Next we integrate the $k$th term by parts $k$ times to get
\begin{equation}
  I_f = \int f(\xi) \, \sum_{k=0}^\infty {(-1)^k \over k!} \, 
        {d^k h^k(\xi) \over d\xi^k} \, d\xi,
\end{equation}
and finally change variables again from $\xi$ to $x \equiv \xi+1$,
\begin{equation}
  I_f = \int_{-\infty}^\infty f(x-1) \, \sum_{k=0}^\infty {(-1)^k \over k!}
        \, {d^k h^k(x-1) \over dx^k} \, dx.
\end{equation}
This result has the same form as the original expression for $I_f$, but
with $d\xi/dx$ replaced by an expansion.  However, since the test function
$f$ is arbitrary, this requires the expansion and $d\xi/dx$ to be equal:
\begin{equation}
  {d\xi \over dx} = \sum_{k=0}^\infty {(-1)^k \over k!}
        \, {d^k h^k(x-1) \over dx^k}.
\end{equation}

We now turn to the second stage of the calculation, namely finding
solutions to the third-order approximation (\ref{wnleqn}) to the
evolution equation.  We assume the solution will have a fundamental
wave number $k$ in the unstable range, with an amplitude of order
$\epsilon$.  The quadratic terms in the evolution equation will then
generate a Fourier component in the solution with wave number $2k$ and
possibly a constant term, and the cubic terms will lead to a component
with wave number $3k$.  Thus we write
\begin{eqnarray}
  h(x,t) \approx \epsilon M(t) \cos(kx-\phi(t)) &+& \epsilon^2 M_0(t)
         + \epsilon^2 M_2(t) \cos[2(kx-\phi(t))+\theta_2(t)] \nonumber \\
         &+& \epsilon^3 M_3(t) \cos[3(kx-\phi(t))+\theta_3(t)] + \cdots,
\label{wnlansatz}
\end{eqnarray}
allowing phase differences among the various Fourier components.  We
substitute this ansatz into the evolution equation and expand in powers
of $\epsilon$.  Then the coefficients of $\cos(kx-\phi(t))$ and
$\sin(kx-\phi(t))$ give equations for the fundamental amplitude $M(t)$
and phase $\phi(t)$:
$$  \dot M = \sigma_k M - \epsilon^2 J k^2 M_0 M \cos k
              - {\epsilon^2 \over 2} J k^2 M_2 M \cos(k-\theta_2)$$
$$             + {\epsilon^2 \over 8} J k^3 M^3 \sin k + O(\epsilon^4) 
\eqno (32)$$
$$  \dot \phi = \omega_k  - \epsilon^2 J k^2 M_0 \sin k
              - {\epsilon^2 \over 2} J k^2 M_2 \sin(k-\theta_2)
              - {\epsilon^2 \over 8} J k^3 M^2 \cos k + O(\epsilon^4).
\eqno (33) $$

Evidently we need to find $M_0$, $M_2$, and $\theta_2$ in order to
determine the amplitude $M$ and propagation velocity $\dot\phi/k$ of the
pattern.  The $x$-independent term in the expansion of the evolution
equation gives $\dot M_0 = 0$, so $M_0$ is in fact a constant; as argued
above, we can choose it to be zero, so the $M_0$ terms in the $M$ and
$\phi$ equations can be dropped.  The equations for $M_2$ and $\theta_2$
come from the coefficients of $\cos(2kx-2\phi+\theta_2)$ and
$\sin(2kx-2\phi+\theta_2)$ in the evolution equation.  These are best
written in the form
$$  \frac{d}{dt} M_2 e^{-i\theta_2} = 
    (\sigma_{2k} + i\omega_{2k} - 2i\omega_k) M_2 e^{-i\theta_2}
               - J k^2 M^2 e^{2ik} + O(\epsilon^2).
\eqno (34)$$
Similarly, we find equations for $M_3$ and $\theta_3$,
$$  \frac{d}{dt} M_3 e^{-i\theta_3} = 
    (\sigma_{3k} + i\omega_{3k} - 3i\omega_k) M_3 e^{-i\theta_3}
               - {9 \over 2} J k^2 M M_2 e^{3ik - i\theta_2}
               - {9 \over 8} i J k^3 M^3 e^{3ik} + O(\epsilon^2).
\eqno (35)$$

Note that in the equation for $\dot M(t)$, all the terms on the right are
of order $\epsilon^2$.  Therefore $M(t)$ changes on a long time scale of
order $\epsilon^{-2}$, while $M_2$ and $\theta_2$ vary on times of order
unity.  Thus we may regard $M$ as a constant in the equations for $M_2$
and $M_3$.  Since $\sigma_{2k}$ is negative, $M_2 \exp(-i\theta_2)$ goes
to a quasi-steady state value which is proportional to $M^2$.
Substituting this value into the $M$ equation gives
$$  \dot M = \sigma_k M - \epsilon^2 \lambda M^3 + O(\epsilon^4) \eqno (36)$$
The analytical expression for the Landau constant $\lambda$ is complicated
and unenlightening; substituting (\ref{Jexpansion}) and (\ref{kexpansion})
into it gives
$$  \lambda = 16.905 + 64.680 \epsilon q + O(\epsilon^2) \eqno (37)$$
Note that the correction terms in the evolution equation for $M$, which
would come from including higher-order terms in the original expansions
(\ref{wnleqn}) for the evolution equation and (\ref{wnlansatz}) for
$h(x,t)$, are of order $\epsilon^4$, not $\epsilon^3$.  As a result, the
$\epsilon^3$ terms in the equation come only from expanding the analytical
expressions for $\sigma$ and $\lambda$ in powers of $\epsilon$.  This also
holds for the equations for the higher harmonics.  Thus we get the
first-order corrections to all of our results essentially for free.

From (37) 
and the third-order expansion (\ref{wnlsigma})
for $\sigma_k$ we obtain the steady-state amplitude $M$ to first order
in $\epsilon$,
$$  M^2 = (0.25946 - 0.59719 q^2) - (0.87725 - 2.10774 q^2) \, \epsilon q
      + O(\epsilon^2) \eqno (38)$$
We then find the phase velocity,
$$  v_{ph} = \omega / k = -1 + 4.4934 \epsilon q + (1.877 - 4.320 q^2) \epsilon^2
           - (3.439 - 10.128 q^2) \epsilon^3 q + O(\epsilon^4),
\eqno (39)$$
and the group velocity,
$$  v_{gr} = d\omega / dk = 19.1907 + 13.4802 \epsilon q + 
    (18.241 - 40.984 q^2) \epsilon^2 - (43.198 - 107.23 q^2) \epsilon^3 q
    + O(\epsilon^4),\eqno (40)$$
from (33).
  The equation (34)
for $M_2$ and $\theta_2$
gives
$$  M_2 / M^2 
=0.908115 + 0.50118 \epsilon q, \nonumber \\$$
$$  \theta_2 / \pi =
0.22916 - 0.36189 \epsilon q,\eqno (41)$$
and from (35)
we find
$$  M_3 / M^3 
=1.52273 + 1.89385 \epsilon q, \nonumber \\$$
$$  \theta_3 / \pi =
0.4309 - 0.6024 \epsilon q \eqno (42)$$

As usual, the ripple solutions we have found are not all stable; instead,
those with too large a wave number deviation $q$ are linearly unstable.
The calculation of the critical value of $q$ is rather intricate, so we
defer it to the Appendix.  The result is that the range of stable wave
numbers is rather wider than usual -- it extends out to $q = 0.9095 q_0$,
where $q_0 = (2/J_c)^{1/2}$ is the wave number deviation at which $\sigma_k$
vanishes to leading order in $\epsilon$.  At the edge of the stable range,
the amplitude $M$ of the ripple solution is 0.4157 times its value ay $q=0$.

If instead of the NO saltation relation we use the symmetric relation
(\ref{KBHxi}), the results of the analysis are rather different.  The
expansion of $d\xi/dx$ is not as simple and clean as the derivation
above; the evolution equation (\ref{wnleqn}) is replaced by
$$ {\partial h(x,t) \over \partial t} =
h''(x,t) - J [h(x-1,t) - h(x,t)]'
  + J \{ [h(x-1,t) - h(x,t)] h'(x-1,t) \}' $$
$$ {J \over 2} \{ [h(x-1,t) - h(x,t)]^2 h''(x-1,t) \}'
   - J \{ [h(x-1,t) - h(x,t)] [h'(x-1,t)]^2 \}' + \cdots \eqno (43)$$
We again substitute the ansatz (\ref{wnlansatz}) into this equation and
work out the Fourier components of the result.  The equations for $M$ and
$\phi$ become
$$  \dot M = \sigma_k M
        - \epsilon^2 J k^2 M_2 M [\cos k \cos \theta_2 - \cos(2k-\theta_2)]$$
$$        + {\epsilon^2 \over 4} J k^3 M^3 (\sin k - 2 \sin 2k) , \eqno (44)$$
$$  \dot \phi = \omega_k
        + \epsilon^2 J k^2 M_2 [\cos k \sin \theta_2 + \sin(2k-\theta_2)]
        - {\epsilon^2 \over 2} J k^3 M^2 (\cos k - \cos 2k) \eqno (45)$$
where now $\omega_k$ is given by $- J \, k \, (1 - \cos k)$ as is
appropriate for this model.  Note that the $M_0$ terms which were present
in (32) and (33)
are absent here; this is because the
new saltation relation respects the symmetry under addition of a constant
to $h$.  The evolution of $M_2$ and $\theta_2$ is now given by
$$  \frac{d}{dt} M_2 e^{-i\theta_2} = 
    (\sigma_{2k} + i\omega_{2k} - 2i\omega_k) M_2 e^{-i\theta_2}
               + J k^2 (e^{ik} - e^{2ik}) M^2 + O(\epsilon^2), \eqno (46)$$
and the third harmonic by
 $$ \frac{d}{dt} M_3 e^{-i\theta_3} = 
    (\sigma_{3k} + i\omega_{3k} - 3i\omega_k) M_3 e^{-i\theta_3}
+ {3 \over 2} J k^2 e^{ik} (1 - e^{ik}) (1 + 3 e^{ik}) 
M M_2 e^{-i\theta_2} $$ 
$$+ {3 \over 8} i J k^3 e^{ik} (1 - e^{ik}) (1 - 3 e^{ik}) M^3 + O(\epsilon^2)
\eqno (47)$$


After carrying out the calculation we find a much larger value for the
Landau coefficient,
$$  \lambda = 151.26 + 88.014 \epsilon q + O(\epsilon^2).\eqno (48)$$
Thus for a given wave number, the restabilized amplitude $M$ of the ripples
is smaller by a factor of about 3:
$$  M^2 = (0.0289975 - 0.066743 q^2) - (0.003966 - 0.0190315 q^2) \, \epsilon q
      + O(\epsilon^2).\eqno (49)$$
The phase velocity is more negative than before, as we found from the
linear stability analysis,
$$  v_{ph} = -5.6033 + 4.4934 \epsilon q - (1.506 - 1.16475 q^2) \epsilon^2
           - (0.197 - 1.853 q^2) \epsilon^3 q + O(\epsilon^4), \eqno (50)$$
and likewise the group velocity,
$$  v_{gr} = 14.5874 + 13.4802 \epsilon q - (2.569 - 4.612 q^2) \epsilon^2
           - (8.152 - 19.799 q^2) \epsilon^3 q + O(\epsilon^4). \eqno (51)$$
For a given amplitude, however, the harmonics are stronger than before:
we find
$$  M_2 / M^2 
=1.41691 + 0.213856 \epsilon q, \nonumber \\$$
$$  \theta_2 / \pi =
x0.4443 - 0.2027 \epsilon q,\eqno (52)$$
and
$$  M_3 / M^3 =
2.89266 + 1.57014 \epsilon q, \nonumber \\$$
$$  \theta_3 / \pi =
-0.1426 - 0.3952 \epsilon q. \eqno (53)$$
The range of wave numbers for which these solutions is stable is somewhat
narrower than before but still wider than usual, extending out to
$q = 0.7571 q_0$, where the amplitude $M$ is 0.6533 times its value at $q=0$.

\section{Numerical solutions}

We now present numerical solutions and compare them with the
predictions made in the previous section.  The nonlocal evolution
equation (\ref{heqn}) was solved numerically with periodic boundary
conditions on a system of length $l = 2\pi/k$, so that only the
Fourier modes $nk$ contributed to the solutions.  For the discretization 
scheme, we chose an explicit method using 
forward differences in time and central differences in space.
The axis was discretized at
$2^9$ equally spaced sites with $\Delta x= 2\pi/k\bullet 2^{-9}$
and solutions were generated for
five different values of $J$ near $J_c=4.603$, namely $J=4.62, 4.65,
4.70, 4.75, 4.80$, and values of $k$ were chosen to span the unstable
region.  Initial conditions were sinusoids of wave number $k$ centered
around $h=0$.
 At t=0, we start with a sinusoid at a particular
wave number $k$, and let it evolve with $(\Delta x)^2/\Delta t=1/4$ 
until it reaches a steady state.  This
takes about $10^6$ time steps.  The nonlocal term in the evolution equation,
$J\left({d\xi}/{dx}-1\right)$, was evaluated for a given $x$ by finding the 
nearest upwind value of $\xi$ satisfying the equation $x=\xi+1+h(\xi)$.
Spefifically, the first root of the function $f(\xi;x) = \xi +
1 + h(\xi) - x$ with value less than $x$ was obtained by simply finding
the two sites upwind of $x$ and nearest to it between which $f(\xi;x)$
changed sign.  Then, ${d\xi}/{dx}$ was calculated using the values of
$h$ at these sites.
The final steady state, $h(x,t)$, is then Fourier transformed, i.e.:
$$  h(x,t) = \sum_{n=1}^\infty [a_n(t) \sin nkx + b_n(t) \cos nkx ],\eqno (54)
$$
from which we obtain
$M_n = (a_n^2 + b_n^2)^{1/2}$ and $\theta_n = \tan^{-1}(b_n/a_n)$.  Note
that $M_1 = M$ and $\theta_1 = \phi$ in this notation.  The nonlinear
analysis in the previous section predicts that these quantities will
go to time-independent values.  We find numerically that they
actually oscillate as a function of time around their mean values.
However, the magnitudes of these oscillations are quite small and
decrease with increasing grid resolution, so we believe them to be
numerical artifacts.  We therefore take the time averaged mean
values and compare them with the predictions of the weakly nonlinear
analysis.

We also find that although we start with an initial profile with the
average height $h_0 = 0$, the mean position of the steady state
pattern shifts slightly upward in some cases, downward in others,
 to a small but finite $h_0$.  Since
the mean height of the sandbed is conserved by the exact evolution
equation, we believe that this is also a numerical artifact.  Moreover,
as mentioned in section 2, we can map any steady state solution with
finite $h_0$ to the solution with $h_0 = 0$ by redefining the
horizontal length scale from $L_0$ to $L_0(1+h_0)$ and shifting the
control parameter from $J$ to $J (1+h_0)$.  However, the magnitude
offset $h_0$ was always of the order of $10^{-5}$ to $10^{-3}$, and
thus in all cases studied here, the corrections due to such an
offset are quite negligible.  Hence, our results without these
corrections are virtually identical to those with corrections. 

Figure 2 shows the amplitude $M$ of the fundamental mode as a
function of $k$ for different values of $J$.  The data points are
fairly close to the values predicted by the first-order expansion (38),
which are represented by continuous curves.  Note that
the curves are asymmetric around the critical value $k_c$ and the asymmetry
becomes more pronounced for larger $J$.  The weakly nonlinear analysis
is capable of predicting this asymmetry only because the order-$\epsilon$
terms are included.

In Figure 3 we plot the phase velocity of the fundamental mode
against $k$ for different values of $J$.  The speed was obtained by 
calculating $\phi(t)$ in the expression $cos(kx-\phi(t))$, which is
proportional to the fundamental mode in the steady state.  The function
$\phi(t)$ was found to be linear in t, so v was calculated as 
$v=d\phi(t)/dt/k$.
The data points are compared against the weakly nonlinear
predictions (solid line) given by (39).
Only for fairly
large $J$, and only near the high-wave-number end of the band of ripple
solutions, does the velocity become positive, that is, in the direction
of the wind.

In Figures 4a and 4b are plotted the ratios of the amplitudes of the
second and third harmonics to the appropriate powers of the
fundamental ampltude, i.e., $R_2 = M_2 / M^2$ and $R_3 = M_3 / M^3$.
The data fit quite well with the theoretical predictions for both
cases, in particular near the onset $k_c= 4.493$, where the
nonlinear analysis is most reliable.  Note that the first-order
terms in the analytical results match the slope of the numerical
results.  The curvature which is evident in the numerical data for
$k$ farther from $k_c$ is apparently a higher-order effect.  Note
that the width of the band of ripple solutions increases with
$\epsilon$, so an appreciably large $\epsilon$ is required to reach
these larger values of $|k-k_c|$.

In Figures 5a and 5b we plot the phase angles $\theta_2$ and
$\theta_3$ against $k$.  The agreement between the simulations and
the weakly nonlinear analysis is again quite strong for $J$ near onset.
The order-$\epsilon$ terms in the analytical results match
the slope of the numerical data.  For higher $J$ we observe a
systematic downward deviation in the numerical results.  The shift
appears to be linear in $J$, and so is second-order in
$\epsilon \equiv (J - J_c)^{1/2}$.

\section{Discussion}

We have carried out numerical and weakly nonlinear analyses of the
Nishimori-Ouchi continuum model [3,11] for windblown sand, and also for a
modification of that model which respects the physical symmetry of the
system under changes of the reference level of height.  Both versions
of the model yield the surprising result that the ripple patterns, which
form when the flat sandbed becomes unstable, drift upwind even as the
sand which forms the ripples is blown downwind.  This drift is found
in the linear stability analysis and persists in the weakly nonlinear
results, and numerical integrations confirm that it is a real consequence
of the model.  Such a counterintuitive result has not been examined or
detected by previous Monte Carlo simulations of this model [3,11] or in 
real experiments [10].  It would be interesting to check experimentally
whether or not ripples can move against the wind.  The symmetric version
of the model actually predicts a considerably higher upwind drift speed
than the original Nishimori-Ouchi version.

It may also be surprising that the differences between the symmetric and
Nishimori-Ouchi models are merely qualitative.  The restabilized ripple
pattern for a given value of the control parameter have smaller amplitudes
(by a factor of about 3) and higher drift velocities (by a factor of over
5) in the symmetric model than in the original version.  The relative
sizes and phases of the higher harmonics in the ripple shape are also
different for the two models.

A number of modifications to the model are needed in order to make it
comparable with experiments.  A major ingredient that is left out of the
model is any effect of the surface topography on the wind.  This
lack means that there is no shadowing effect in the model.  Including
such an effect would make it more likely for grains to settle on the
downwind side of a ripple than on the upwind side, and more likely for
them to be blown off the upwind side than the downwind side.  This
would likely reduce the tendency of the ripples to drift upwind.  The
result that the ripples drift upwind in this model, which neglects
shadowing, may be an indirect indication of the importance of shadowing in
the development of real ripple patterns.  An improved model of creep
may also be needed; a downwind bias in the creep would modify the
drift velocity.  Perhaps most critical is a better and more realistic
form of the saltation function, which must account the effects of the
topography of the sandbed, and the many particle dynamics of the grains
in the air as well as on the surface.

\newpage

\noindent {\bf References}
\vskip 0.2 true cm

\noindent [1] R.A. Bagnold, Proc. Roy. Soc. A157, 594 (1936);
See also {\it The physics of blown sand and desert dunes},
reprinted by Chapman and Hall (1981).

\noindent [2] K. Pye and H. Tsoar, {\it Aeolian sand and sand dunes},
Unwin Hyman, London (1990).

\noindent [3] H. Nishimori and N. Ouchi, Phys. Rev. Lett. {\bf 71}, 197 (1993).

\noindent [4] R.S. Anderson, Sedimentology, {\bf 34}, 943 (1987).

\noindent [5] O. Terzidis, P. Claudin, and J.P. Bouchaud, cond-matt/9801295.

\noindent [6] For recent observations of surface instabilities that
develop when grains are subject to vibrations, see:
F. Melo, P. Umbanhowar, and H.L. Swinney, Phys. Rev. Lett.
{\bf 75}, 3838 (1995); T.H. Metcalf, J.B. Knight, and H.M. Jaeger, 
Physica A {\bf 236}, 202 (1997); K.M. Aoki and T. Akiyama, 
Phys. Rev. Lett. {\bf 77}, 4166 (1996).

\noindent [7] See, e.g.: J.S. Langer, Science {\bf 243}, 1150 (1989);
D.A. Kessler, H. Levine, and J. Koplik, Adv. Phys. {\bf 37}, 255 (1988);
{\it Dynamics of curved fronts}, edited by P. Pelce, 
Academic, San Diego, (1988) and references therein.

\noindent [8] See, e.g., M. Ben-Amar and B. Moussallam, Phys. Rev. Lett.
{\bf 60}, 317 (1989) and references therein.

\noindent [9] D. A. Kurtze and D. C. Hong, J. Korean. Phys. Soc.
{\bf 28}(2), 178 (1995) and references therein.

\noindent [10] A. Betat, V. Frette, and I. Rehberg, Phys. Rev. Lett.
{\bf 83}, 88 (1999).

\noindent [11] More detailed description of the model can be found in:
H. Nishimori and N. Ouchi, Int. J. Mod. Phys. B. Vol.7, \# 9 \& 10, 2025 
(1995).

\newpage

\appendix
\section {Stability of ripple solutions and the Eckhaus boundary}

In this section, we examine the stability of the ripple solutions, and so
determine the Eckhaus boundary in the $J$--$k$ plane, within which the
solutions are linearly stable and so may be observed.  We begin with the
ripple solution
$$  h_0(x,t) = \epsilon M \cos(kx - \omega t)
          + \epsilon^2 M_2 \cos[2(kx - \omega t) + \theta_2]
          + O(\epsilon^3 M_3), \eqno (A-1)$$
with $k = k_c + \epsilon q$, and add an infinitesimal perturbation
$h_1(x,t)$.  If the perturbation contains a Fourier component with wave
number $k + \epsilon q'$, then the nonlinear terms in the evolution equation
will generate a component with wave number $k - \epsilon q'$.  Thus we will
start the calculation by taking $h_1$ to have the form
$$  h_1(x,t) = A_-(t) \cos[(k - \epsilon q')x - (\omega t + \phi_-(t))]
           + A_+(t) \cos[(k + \epsilon q')x - (\omega t + \phi_+(t))].
\eqno (A-2)$$
Substituting this into the evolution equation, expanding, and picking
off the coefficients of the sines and cosines of $(k - \epsilon q')x$
and $(k + \epsilon q')x$ yields a closed set of equations for the
amplitudes $A_-$ and $A_+$ and the phase $\phi_+ + \phi_-$.  These
equations have the form
$$  \dot A_- = 
\Sigma_- A_- + \alpha A_+ \cos\psi,$$ 
$$  \dot A_+ =
\Sigma_+ A_+ + \alpha A_- \cos\psi,$$ 
$$  \dot \psi =
\Omega - \alpha [(A_+/A_-) + (A_-/A_+)] \sin\psi \eqno (A-3)$$
where $\psi$ is $\phi_+ + \phi_-$ plus a constant which depends on $k$
and $q'$ (but not time), and the overdot denotes a derivative with respect
to the slow time variable $\epsilon^2 t$.  The coefficients are given to
leading order in $\epsilon$ by
$$  \Sigma_\pm =
\frac{k_c^2}{2} \, \left[ \left( \frac{k_c^2}{2} -\frac{2\lambda}{k_c^2} 
\right) \, M^2 \mp 2 q q' - q'^2 \right], \nonumber \\$$
$$  \alpha =
\alpha_0 \, k_c^2 \, M^2 = 1.98183 \, k_c^2 \, M^2, \nonumber \\$$
$$  \Omega =
\frac{1}{2} \, k_c^3 \, M^2 + 3 \, k_c \, q'^2.\eqno (A-4)$$
Note that it is important to keep the second-order term in $h_0$
during the calculation, since it contributes to $\alpha$.  (Omitting
it changes $\alpha_0$ to 2.58559, a 30\% change.)

We must now determine whether the amplitudes given by (A-3)
grow or decay with time.  We can simplify the equations somewhat by
defining
$$  R = A_+ / A_-; \eqno (A-5)$$
the amplitude equations then become
 $$ \dot A_- =
(\Sigma_- + \alpha R \cos\psi) A_+, \nonumber \\$$
$$  \dot R =
(\Sigma_+ - \Sigma_-) R + \alpha (1 - R^2) \cos\psi, \nonumber \\$$
$$  \dot \psi =
\Omega - \alpha [(1 + R^2) / R] \sin\psi.\eqno (A-6)$$
Note that the first equation decouples from the last two.  If it happens
that $R$ and $\psi$ go to constants as $t \to \infty$, then the
amplitudes decay for $\Sigma_- + \alpha R \cos\psi < 0$ and grow
otherwise.  Thus for a given $q$, the ripple state (A-1)
is linearly stable if this inequality is satisfied for {\it every\/}
$q'$, otherwise it is unstable.  To see what $R$ and $\psi$ actually
do, we combine the $R$ and $\psi$ equations into an evolution equation
for the complex variable
$$  Z = R \exp(i\psi),\eqno (A-7)$$
namely
$$  \dot Z = \alpha (1 - Z^2) + (\Sigma_+ - \Sigma_- + i \Omega) Z.\eqno 
(A-8)$$
Clearly this equation has two fixed points, and solving it exactly
reveals that the one with positive real part is a {\it global\/}
attractor and the one with negative real part a global repeller.
Thus $R \cos\psi$ does go to a constant, and from its value we can
decide whether the ripple state is stable or not.

Since it is the real part of $Z$, namely $R \cos\psi$, which determines
whether the perturbation grows or decays, it is useful to rewrite
(A-8) in terms of the real and imaginary parts of $Z$,
$$  Z = X + iY. \eqno (A-9)$$
We find
$$  \dot X =
\alpha (1 - X^2) + (\Sigma_+ - \Sigma_-) X 
           - \Omega Y + \alpha Y^2, \nonumber \\$$
$$  \dot Y =
(\Sigma_+ - \Sigma_-) Y + \Omega X - 2 \alpha X Y.\eqno (A-10)$$
By linearizing about a fixed point $(X,Y)$ of this system, we 
quickly find that the fixed point is an attractor for
$X > (\Sigma_+ - \Sigma_-) / 2 \alpha$.  To find the fixed points,
we set $\dot X = \dot Y = 0$ and solve the second equation for
$Y$ in terms of $X$, then substitute into the first equation to get
$$  \alpha - X (\alpha X - \Sigma_+ + \Sigma_-) =
           \frac{\Omega^2 X (\alpha X - \Sigma_+ + \Sigma_-)}
             {(2 \alpha X - \Sigma_+ + \Sigma_-)^2}. \eqno (A-11)$$
The two sides of this equation are plotted in Fig. 6.  Both sides
are symmetric about $X = (\Sigma_+ - \Sigma_-) / 2 \alpha$, so there is
clearly one solution with $X$ greater than this -- the attractor --
and one, the repeller, with $X$ less.  In order for the perturbation
to decay, the attractor must have $X$ less than $-\Sigma_- / \alpha$.
From the plot, we see that this means that at $X = -\Sigma_- / \alpha$
the right side of (A-11)
must be greater than the left side.
After a little algebra, we can write this condition for the perturbation
to decay in the form
$$  \Sigma_+ \Sigma_- > {\alpha^2 (\Sigma_+ + \Sigma_-)^2 \over
                       (\Sigma_+ + \Sigma_-)^2 + \Omega^2}.\eqno (A-12)$$

Equation (A-12)
above is the condition for the amplitude
of a perturbation with a {\it specific\/} value of $q'$ to decay.
In order to conclude that the ripple solution with a given $q$ is
linearly stable, we must see to it that this condition is satisfied
for {\it all\/} $q'$.  For this we must substitute for the parameters
from (A-4)
 above.  To put the result into a useful form,
we define $Q = 2q'^2 / k_c^2 M^2$ and eliminate $q^2$ in favor of $M^2$.
After some rearranging, we find that the condition for the solution to
be stable is
$$  M^2 > \frac{16 Q}{J} \frac{k_c^2 (\beta+Q)^2 + (1+3Q)^2}
        {k_c^2 [(\beta-Q)^2 + 4Q] [k_c^2 (\beta+Q)^2 + (1+3Q)^2] -
         16 \alpha_0^2 (\beta+Q)^2}, \eqno (A-13)$$
where $\beta = 1 - (4\lambda/k_c^2) = 0.834132$.  The complicated
function of $Q$ on the right has a single maximum for positive $Q$,
at a height of 0.04484.  Ripple states with $M^2$ below this are 
unstable, while those with $M^2$ larger than this are linearly stable.
From this we find that the range of wave numbers of linearly stable
ripple solutions is given by $|q| < 0.9095 q_0$, where
$q_0 = (2/J_c)^{1/2}$ is the largest wave number for which a ripple
solution exists.

In summary, we have found that in the weakly nonlinear regime, the
flat sandbed is unstable against perturbations with wave numbers $k$
in the range
$$  |k - k_c| < \epsilon q_0 = 
    \sqrt{ 2 \left( \frac{J}{J_c} - 1 \right) }, \eqno (A-14)$$
while the Eckahus boundary is given by
$$  |k - k_c| < 0.9095 \epsilon q_0 =
    \sqrt{ 1.654 \left( \frac{J}{J_c} - 1 \right) }. \eqno (A-15)$$

For the symmetric saltation relation, the structure of the calculation
is the same but the numbers are different.  We find that the marginally
stable wave number is given by $q = 0.7571 q_0$, so the Eckhaus boundary
is now given by
$$  |k - k_c| < 0.7571 \epsilon q_0 =
    \sqrt{ 1.1465 \left( \frac{J}{J_c} - 1 \right) }. \eqno (A-16)$$

\newpage

\noindent {\bf Figure Captions}
\vskip 1.0 true cm

\noindent Fig. 1:
Saltation refers to the process of a single grain being ejected from the
surface at a point $\xi$ and being blown to a landing point $x$ by the wind.

\noindent Fig. 2:
The steady state amplitude $\epsilon M$ vs. $k$ for five different values
of $J$.  The continuous lines are the analytical predictions given by
Eq. (38).

\noindent Fig. 3:
The propagation speed of the steady state patterns $v_{ph}$ vs. $k$ for
different values of $J$.   Numerically obtained values are compared to
the analytical predictions (continuous curves) from Eq. (39).

\noindent Fig. 4:
The ratios (a) $R_2 = M_2 / M^2$ and (b) $R_3 = M_3 / M^3$ are plotted
against $k$.  The solid line is the analytical predictions from
Equations (41) and (42).

\noindent Fig. 5:
(a) $\theta_2$ and (b) $\theta_3$ are plotted against $k$ for five
different values of $J$.  The solid line is the analytical predictions
from Equations (41) and (42).

\noindent Fig. 6:
The right and left hand sides of Eq. (A-11).

\newpage

\centerline{\hbox{
\psfig{figure=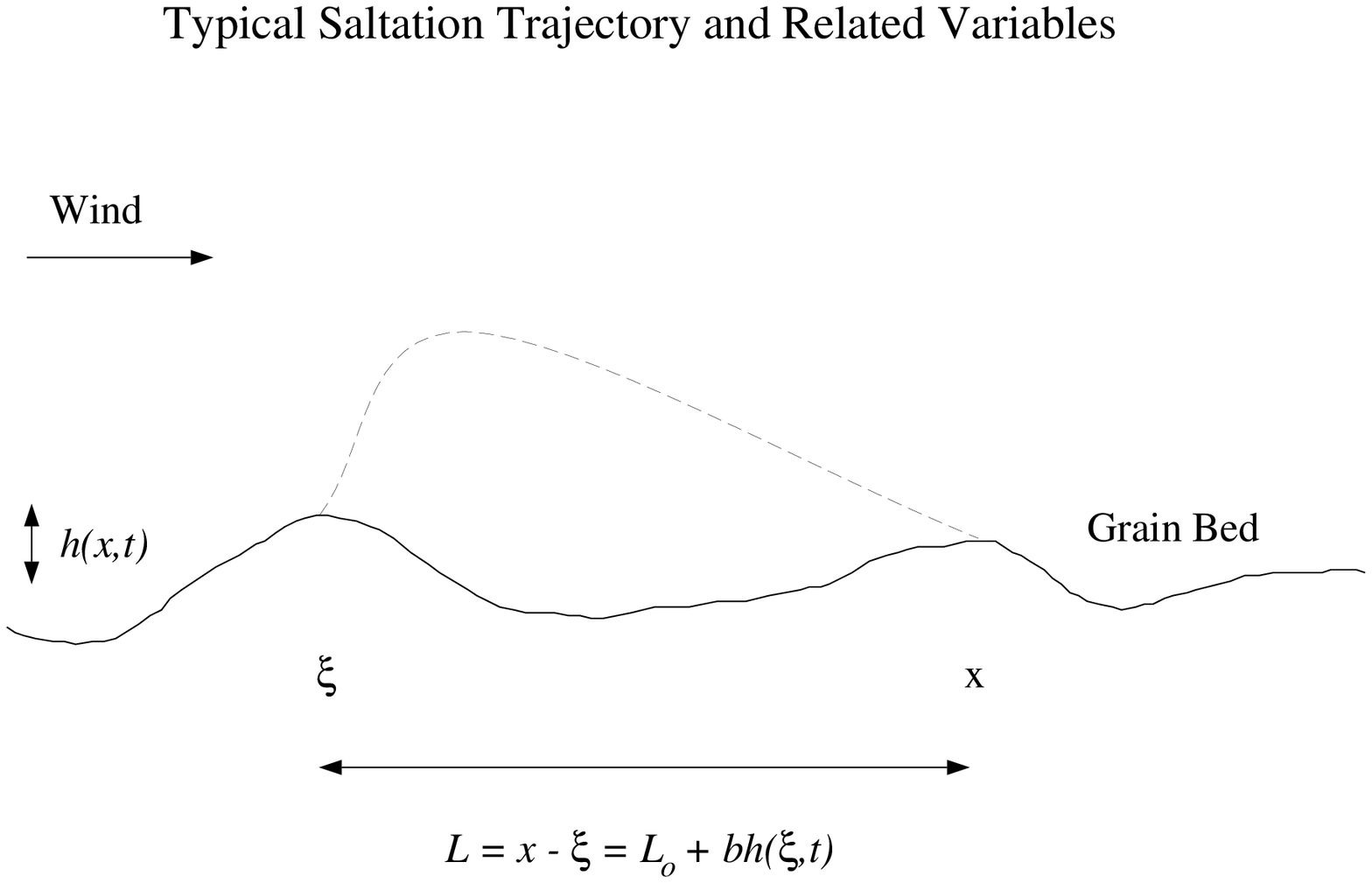}
}}

\thispagestyle{empty}
\centerline{\hbox{
\psfig{figure=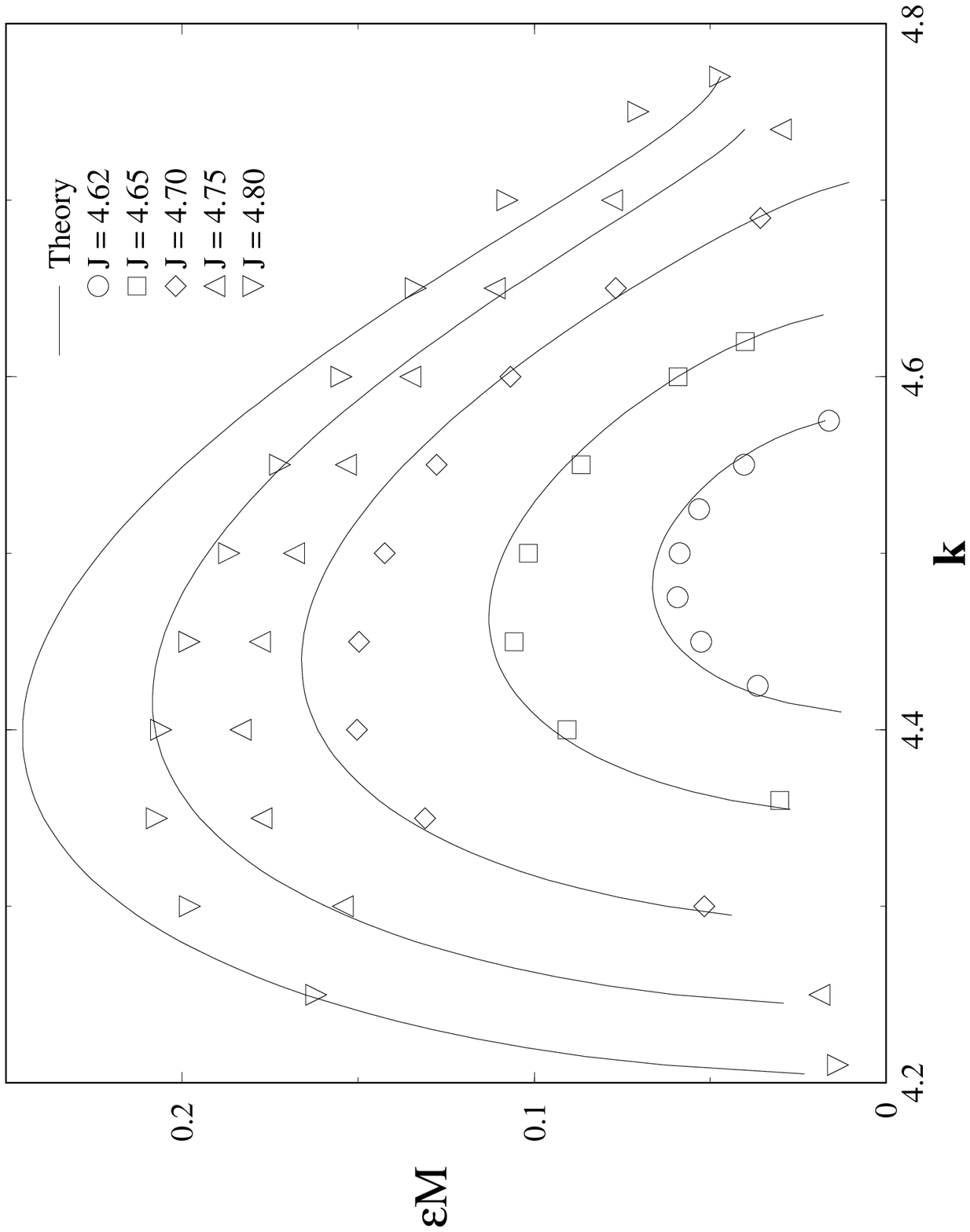}
}}

\thispagestyle{empty}
\centerline{\hbox{
\psfig{figure=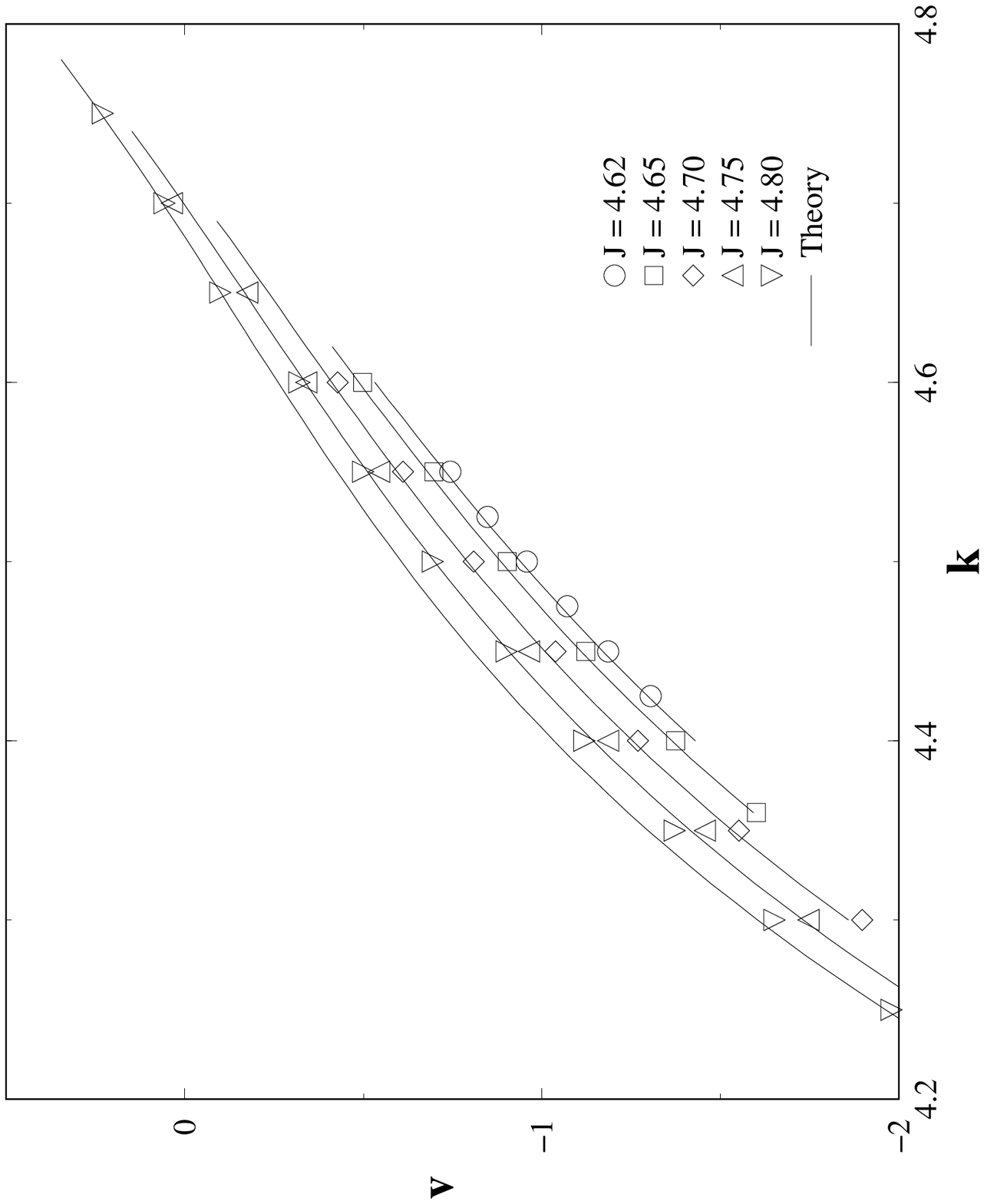}
}}

\centerline{\hbox{
\psfig{figure=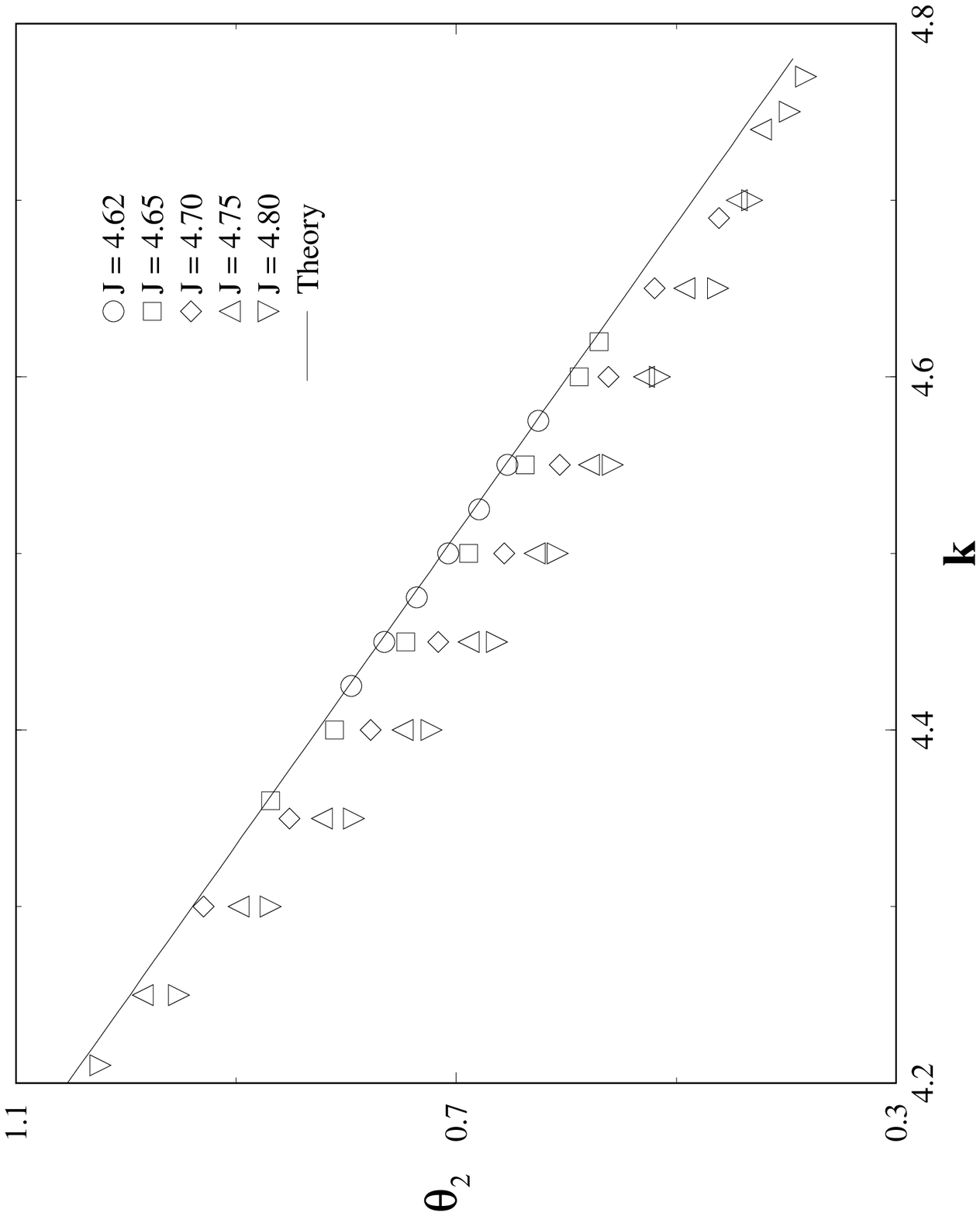}
}}

\centerline{\hbox{
\psfig{figure=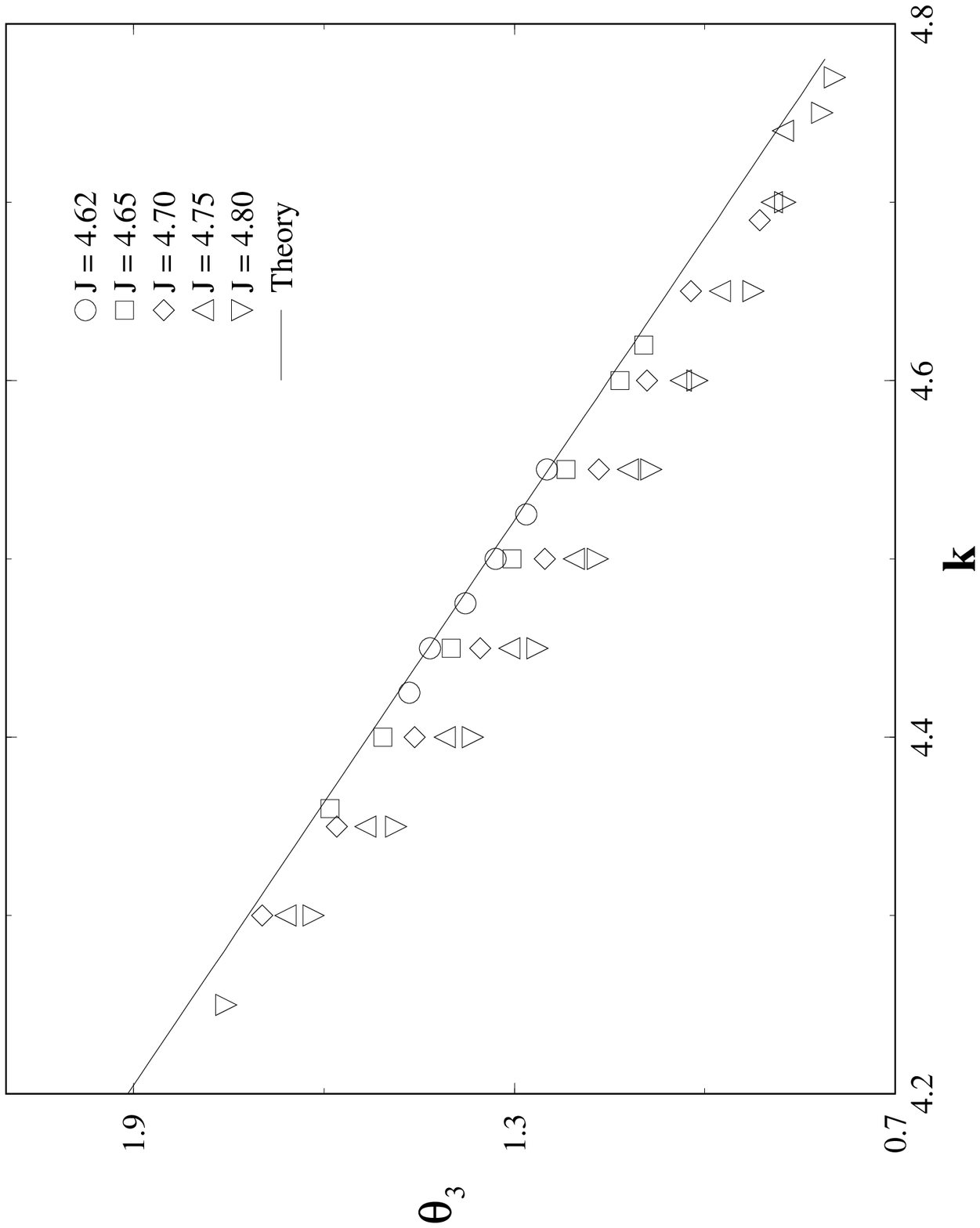}
}}

\centerline{\hbox{
\psfig{figure=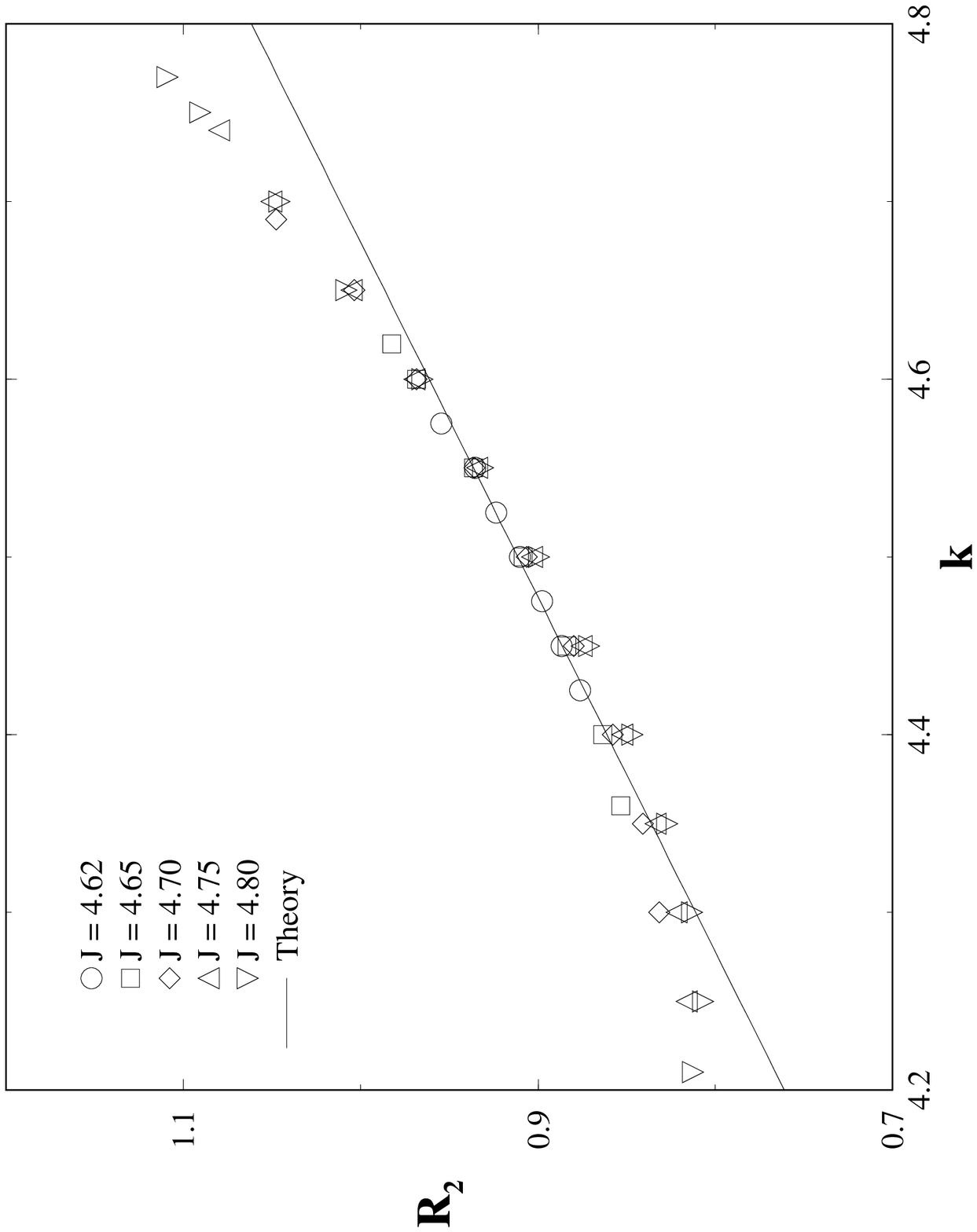}
}}

\centerline{\hbox{
\psfig{figure=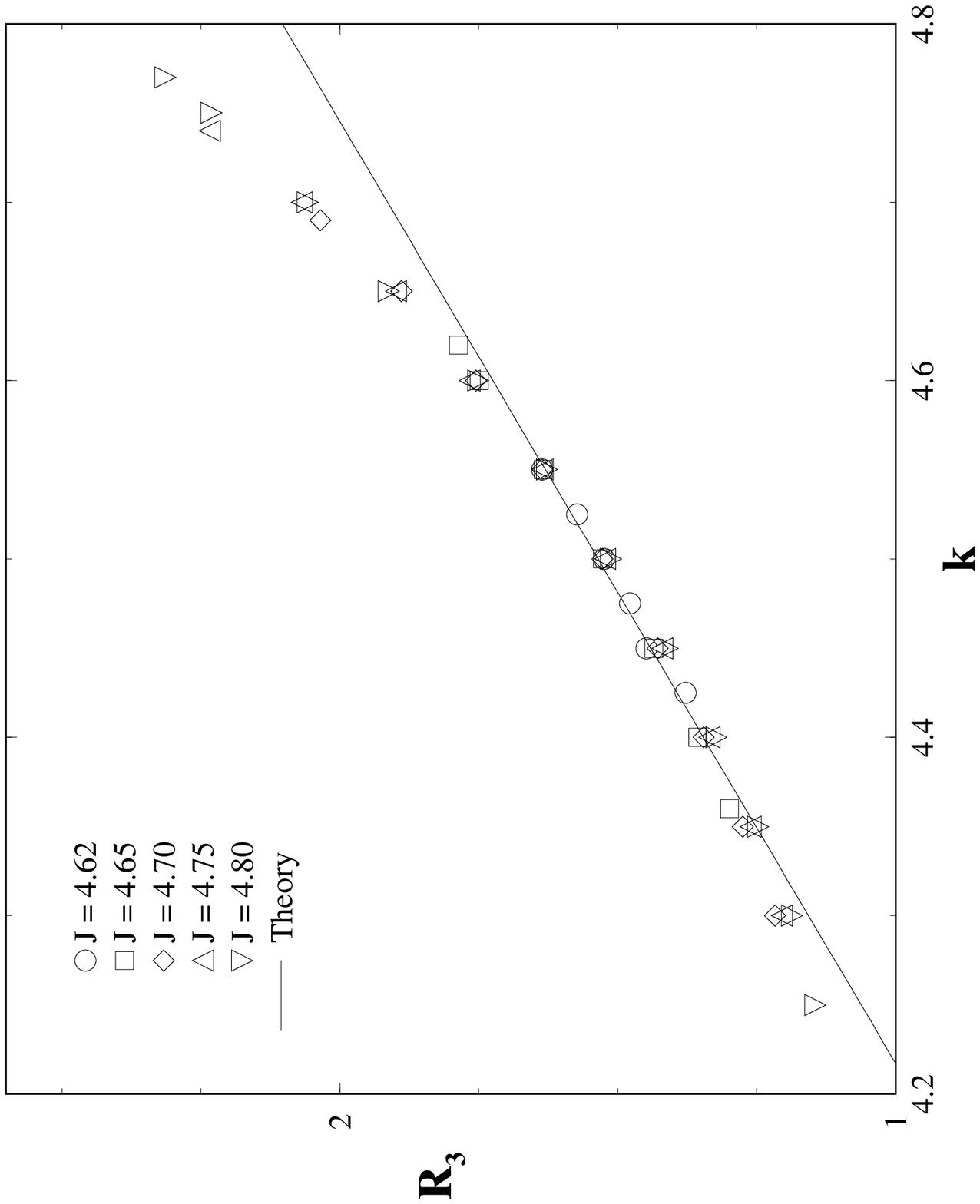}
}} 

\centerline{\hbox{
\psfig{figure=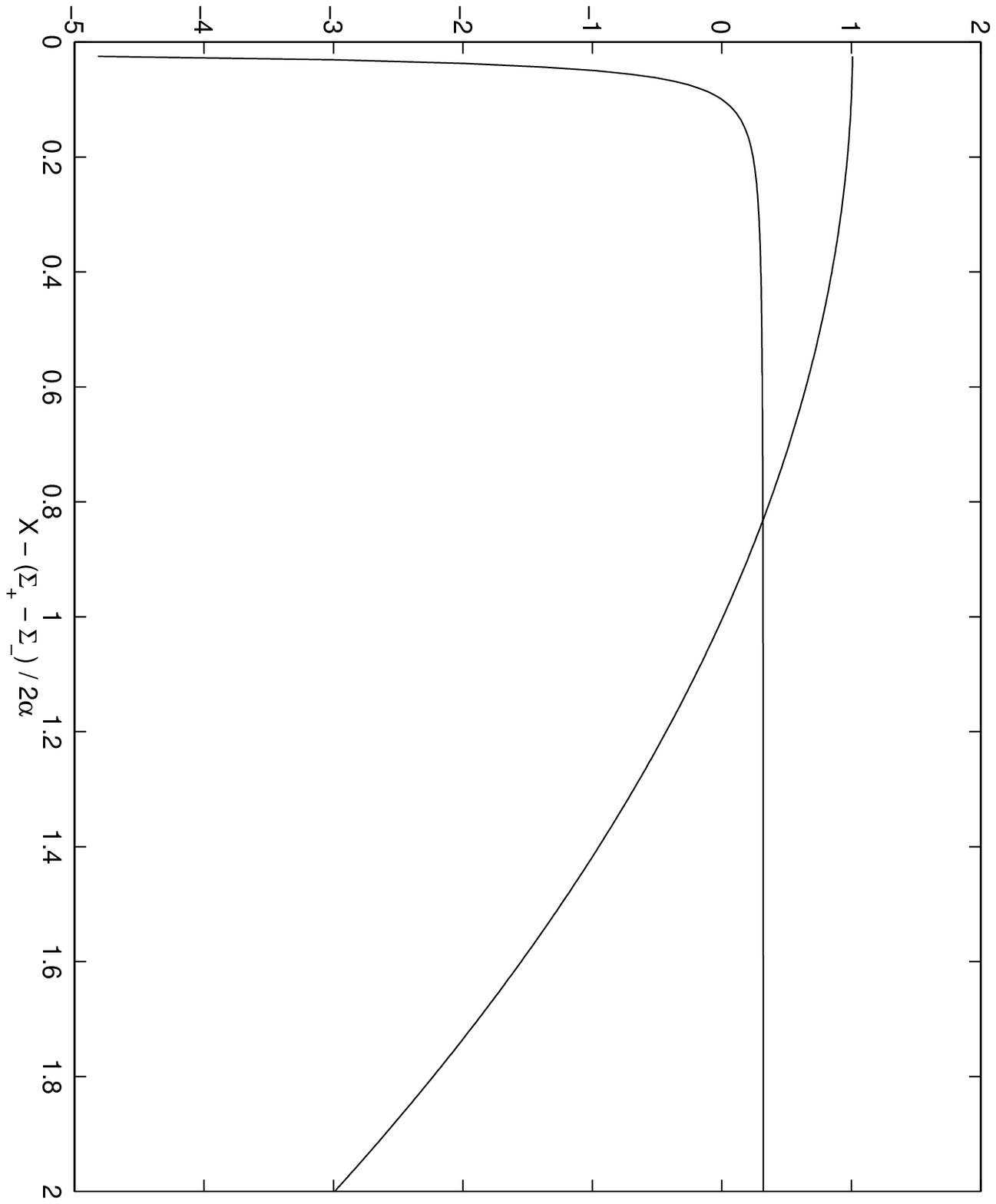}
}}


\end{document}